\documentclass{zermatt}
\usepackage{graphicx}
\usepackage{natbib}
\usepackage[T1]{fontenc}
\usepackage[utf8]{inputenc}
\usepackage[english]{babel}
\usepackage[scaled]{helvet}
\newcommand{\CI}{[C\,{\sc i}]}

\newcommand{\simgt}{\lower.5ex\hbox{$\; \buildrel > \over \sim \;$}}
\newcommand{\simlt}{\lower.5ex\hbox{$\; \buildrel < \over \sim \;$}}
\begin{document}
\title{Unbiased surveys of dust-enshrouded galaxies using ALMA} 
\runningtitle{ALMA Lensing Cluster Survey}
\author{K.~Kohno}
\address{Institute of Astronomy, Graduate School of Science, The University of Tokyo, 2-21-1 Osawa, Mitaka, Tokyo,\\ \email{kkohno@ioa.s.u-tokyo.ac.jp}}
\author{S.~Fujimoto}
\address{Department of Astronomy, The University of Texas at Austin}
\secondaddress{Cosmic Dawn Center (DAWN)}
\author{A.~Tsujita}
\sameaddress{1}
\author{V.~Kokorev}
\address{Kapteyn Astronomical Institute, University of Groningen}
\author{G.~Brammer}
\sameaddress{3}
\secondaddress{Niels Bohr Institute, University of Copenhagen}
\author{G.~E.~Magdis}
\sameaddress{3,~5}
\author{F.~Valentino}
\sameaddress{3}
\secondaddress{European Southern Observatory}
\author{N.~Laporte}
\address{Kavli Institute for Cosmology, University of Cambridge}
\secondaddress{Cavendish Laboratory, University of Cambridge}
\author{Fengwu Sun}
\address{Steward Observatory, University of Arizona}
\author{E.~Egami}
\sameaddress{9}
\author{F.~E.~Bauer}
\address{Instituto de Astrof{\'{\i}}sica, Facultad de F{\'{i}}sica, Pontificia Universidad Cat{\'{o}}lica de Chile}
\author{A.~Guerrero}
\address{Departamento de Astronom\'ia, Universidad de Concepci\'on}
\author{N.~Nagar}
\sameaddress{11}
\author{K.~I.~Caputi}
\sameaddress{4}
\author{G.~B.~Caminha}
\address{Max-Planck-Institut f{\"u}r Astrophysik}
\author{J.-B.~Jolly}
\address{Max-Planck-Institut f{\"u}r extraterrestrische Physik}
\author{K.~K.~Knudsen}
\address{Department of Space, Earth and Environment, Chalmers University of Technology, Onsala Space Observatory}
\author{R.~Uematsu}
\address{Department of Astronomy, Kyoto University}
\author{Y.~Ueda}
\sameaddress{15}
\author{M.~Oguri}
\address{Center for Frontier Science, Chiba University}
\author{A.~Zitrin}
\address{Physics Department, Ben-Gurion, University of the Negev}
\author{M.~Ouchi}
\address{National Astronomical Observatory of Japan}
\secondaddress{Institute for Cosmic Ray Research, The University of Tokyo}
\author{Y.~Ono}
\sameaddress{19}
\author{J.~Gonz\'alez-L\'opez}
\address{N\'ucleo de Astronom\'ia de la Facultad de Ingenier\'ia y Ciencias, Universidad Diego Portales}
\secondaddress{Las Campanas Observatory, Carnegie Institution of Washington}
\author{J.~Richard}
\address{Univ Lyon, Univ Lyon1, Ens de Lyon, CNRS, Centre de Recherche Astrophysique de Lyon}
\author{I.~Smail}
\address{Centre for Extragalactic Astronomy, Department of Physics, Durham University}
\author{D.~Coe}
\address{Space Telescope Science Institute}
\author{M.~Postman}
\sameaddress{24}
\author{L.~Bradley}
\sameaddress{24}
\author{A.~M.~Koekemoer}
\sameaddress{24}
\author{A.~M.~Mu\~noz Arancibia}
\address{Millennium Institute of Astrophysics (MAS); Center for Mathematical Modeling (CMM), Universidad de Chile}
\author{M.~Dessauges-Zavadsky}
\address{Observatoire de Gen\'{e}ve, Universit\'{e} de Gen\'{e}ve}
\author{D.~Espada}
\address{Departamento de F\'{i}sica Te\'{o}rica y del Cosmos, Campus de Fuentenueva, Edificio Mecenas, Universidad de Granada}
\author{H.~Umehata}
\address{Institute for Advanced Research, Nagoya University}
\author{B.~Hatsukade}
\sameaddress{1}
\author{F.~Egusa}
\sameaddress{1}
\author{K.~Shimasaku}
\address{Department of Astronomy, Graduate School of Science, The University of Tokyo}
\author{K.~Matsui-Morokuma}
\address{University of Tsukuba}
\author{W.-H.~Wang}
\address{Institute of Astronomy and Astrophysics, Academia Sinica}
\author{T.~Wang}
\address{Key Laboratory of Modern Astronomy and Astrophysics, Nanjing University}
\author{Y.~Ao}
\address{Purple Mountain Observatory and Key Laboratory for Radio Astronomy, Chinese Academy of Sciences}
\author{A.~J.~Baker}
\address{Department of Physics and Astronomy, Rutgers, the State University of New Jersey}
\author{Minju~M.~Lee}
\sameaddress{3}
\author{C.~del P.~Lagos}
\address{International Centre for Radio Astronomy Research (ICRAR), M468, University of Western Australia}
\author{D.~H.~Hughes}
\address{Instituto Nacional de Astrof{\'i}sica, {\'O}ptica y Electr{\'o}nica (INAOE)}
\author{ALCS collaboration}
\begin{abstract}
The ALMA lensing cluster survey (ALCS) is a 96-hr large program dedicated to uncovering and characterizing intrinsically faint continuum sources and line emitters with the assistance of gravitational lensing. 
All 33 cluster fields were selected from \textit{HST}/\textit{Spitzer} treasury programs including CLASH, Hubble Frontier Fields, and RELICS, which also have \textit{Herschel} and \textit{Chandra} coverages. 
The total sky area surveyed reaches $\sim$133 arcmin$^2$ down to a depth of $\sim$60 $\mu$Jy beam$^{-1}$ (1$\sigma$) at 1.2 mm, yielding 141 secure blind detections of continuum sources and additional 39 sources aided by priors. 
We present scientific motivation, survey design, the status of spectroscopy follow-up observations, and number counts down to $\sim$7 $\mu$Jy. Synergies with \textit{JWST} are also discussed. 
\end{abstract}
\maketitle
\section{Introduction}
Long after the pionnering submillimeter (submm) surveys of lensing clusters using SCUBA (e.g., \citealt{Smail1997,Knudsen2008}), 
the advent of ALMA has enabled us to uncover faint ($S_{\rm 1.2mm}\simlt1$ mJy) (sub)mm-selected galaxies, 
which are significantly (up to $\sim$100$\times$) fainter than ``classical'' submm galaxies (SMGs, 
$S_{870\mu m}{\simgt}$\,a few mJy), by individual pointed observations (e.g., \citealt{Hatsukade2013,Fujimoto2016,Gruppioni2020})
including ALMA calibrators \citep{Chen2023} and unbiased deep surveys of a contiguous region like HUDF/GOODS-S
(e.g., \citealt{Aravena2016,Aravena2020,Dunlop2017,Hatsukade2018,Franco2018,Gonzalez-Lopez2017,Gonzalez-Lopez2020}).
%
These faint submm sources are the major contributor to the cosmic infrared background (CIB) light (e.g., \citealt{Fujimoto2016}), in contrast to the classical bright SMGs. 
However, the resolved fraction of CIB remains controversial (e.g., \citealt{MunozArancibia2018}). Furthermore, a fraction of faint ALMA sources are invisible even in the deepest NIR images using \textit{HST} 
(e.g., \citealt{Simpson2014,Fujimoto2016,Gonzalez-Lopez2017,Cowie2018,Schreiber2018,Franco2018,Yamaguchi2019,Casey2019,Williams2019,Umehata2020,Gruppioni2020,Manning2022,Shu2022}).
Recent studies point to the importance of $H$-band-dark but IRAC-detected (a.k.a.\ {\textit{H}-dropout}) galaxies as a key tracer of the early phases of massive galaxy formation, which are not captured by the Lyman break technique relying on the rest-frame UV light (e.g., \citealt{Caputi2014,Wang2019}).
\cite{Smail2021} revealed that SMGs tend to have compact dust continuum size, and more obscured sources tend to exhibit higher star formation rate surface density $\Sigma_{\rm SFR}$, claiming that the extreme, optically dark SMGs are forming spheroids at high redshifts.
Despite this importance, the difficulties to obtain accurate (i.e., spectroscopic) redshifts of these \textit{H}-dropout faint ALMA galaxies (either optical/near-IR spectroscopy or mm/submm line scans) hamper the efforts to characterize them. 
These facts therefore strongly motivate us to search for gravitationally magnified (but intrinsically faint) \textit{H}-dropout ALMA sources. 

\section{ALMA Lensing Cluster Survey}

\begin{figure}[htp]
      \begin{minipage}{0.55\textwidth}
	\includegraphics[angle=0.0,width=\textwidth]{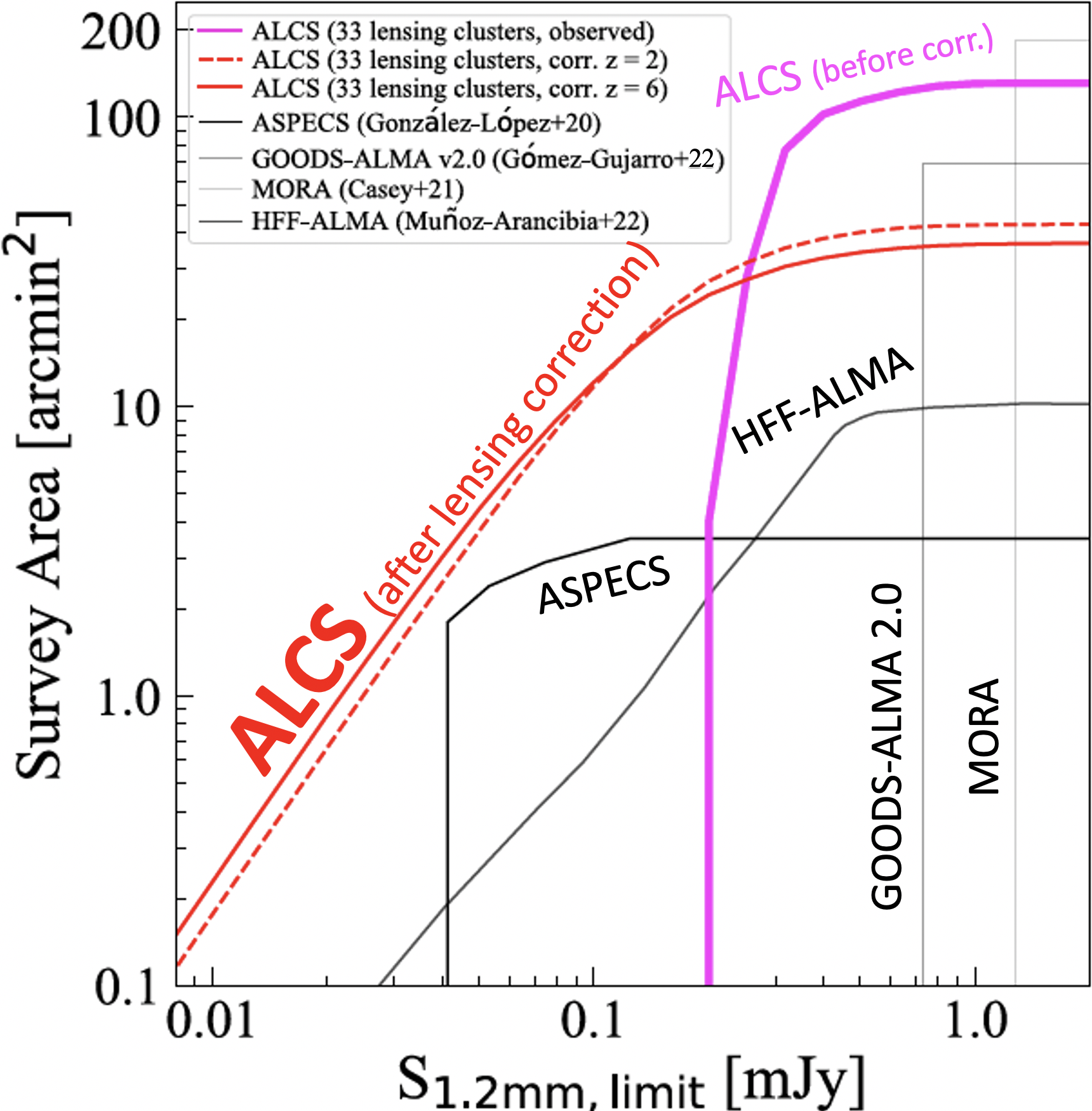}
      \end{minipage}
      \hspace{0.5cm}
      \begin{minipage}[htb]{0.4\textwidth}
	\caption{
Survey areas of ALCS (SNR$>$4.0, the primary beam response of $\simgt$30\%) and other large ALMA surveys in the literature
\citep{Gonzalez-Lopez2020,Gomez-Guijarro2022,Casey2021,MunozArancibia2022}.
The magenta
line denotes the survey area before the lensing correction.
The red dashed and solid lines represent the effective
survey area after the lensing correction, assuming the redshift
at $z = 2.0$ and $z = 6.0$, respectively, with our fiducial
lens models. ALCS explores the unique parameter space towards
faint and wide regimes, compared to previous ALMA
surveys, by exploiting a natural telescope in space, 
i.e., cluster lensing.
 }
	\label{Kohno:fig:SurveyArea}
      \end{minipage}
\end{figure}

\begin{figure}[htp]
\centering
\includegraphics[angle=0.0,width=0.95\textwidth]{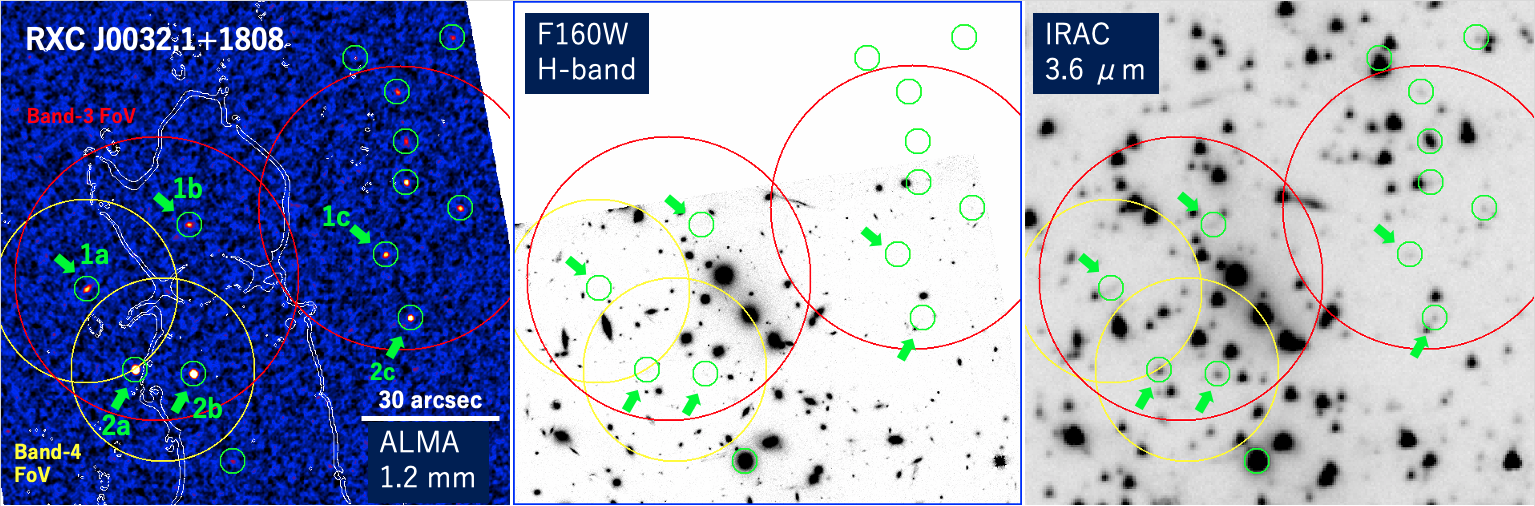}
\caption{ALCS 1.2 mm continuum (left), \textit{HST}/WFC3 \textit{H}-band (middle), and \textit{Spitzer}/IRAC 3.6 $\mu$m images of RXC J0032.1+1808. Two sets of triple image \textit{H}-dropout sources (1a/1b/1c and 2a/2b/2c) are identified. The analysis of ALMA Band-3/4 spectral scans (red and yellow circles) will be presented elesewhere (Tsujita et al., in prep.).
}
\label{Kohno:fig:R0032}
\end{figure}

The base sample of clusters is defined as follows. 
(1) observed by \textit{Planck}, i.e., SZE-based mass measurements are available, 
(2) $z > 0.1$, 
(3) observed by \textit{HST}/ACS with three or more filters, (4) and at least two orbits WFC3/IR (equivalent to the depth of the HST treasury program CANDELS; \citealt{Grogin2011,Koekemoer2011}) with four or more filters, 
(5) also observed by \textit{Spitzer}/IRAC, and 
(6) in the declination range suited for ALMA, i.e., 
$-75^{\circ} < \delta < +25^{\circ}$, to ensure shadowing of less than 5\%. The samples are selected from the best-studied clusters drawn from Cluster Lensing And Supernova Survey with Hubble (CLASH; \citealt{Postman2012}), Hubble Frontier Fields (HFF; \citealt{Lotz2017}), and the Reionization Lensing Cluster Survey (RELICS; \citealt{Coe2019}). We find that 60 clusters fulfill the criteria, and for ALMA cycle-6 we select 33 clusters with the most accurate mass models determined so far, after excluding clusters in the ALMA archive.
The total sky area surveyed reaches $\sim$133 arcmin$^2$ down to a depth of $\sim$60 $\mu$Jy beam$^{-1}$
(1$\sigma$) at 1.2 mm. Fig.~\ref{Kohno:fig:SurveyArea} gives the survey area of ALCS, along with other ALMA surveys. It yields 141 secure blind detections of continuum sources and additional 39 sources aided by priors (Fujimoto et al.\ in prep.).
Fig.\ref{Kohno:fig:R0032} presents an example of the ALCS fields, RXC J0032.1+1808. It shows rich ALMA 1.2 mm sources, including known multiply imaged dusty galaxy at $z=3.631$ (\citealt{Dessauges-Zavadsky2017}, not visible in the Figure). Two sets of triple-image \textit{H}-dropout sources have been uncovered (\citealt{Sun2022}, Tsujita et al., in prep.).
An association of a highly magnified ($\mu\sim$10) \textit{H}-dropout ALMA source with a MUSE-selected galaxy group at $z=4.32$ behind ``El Gordo'' cluster suggests that such dust-enshrouded ALMA sources  can be signposts of richer star-forming environments at high redshifts \citep{Caputi2021}. 
Multi-wavelength mosaics across all 33 ALCS fields and photometric catalogs by reprocessing archival data from \textit{Hubble} \& \textit{Spitzer} have been created \citep{Kokorev2022}. This rich source catalog has been used to conduct stacking analysis (e.g., \citealt{Jolly2021}).
Joint analysis with \textit{Herschel} has been conducted to investigate far-infrared SEDs of ALCS sources \citep{Sun2022}. 
A spin-out program to search for lensed \textit{H}-dropout/faint IRAC sources has been launched \citep{Sun2021}. \textit{Chandra} X-ray properties of ALCS sources have been investigated \citep{Uematsu2023}.

\begin{figure}[htp]
\centering
\includegraphics[angle=0.0,width=0.99\textwidth]{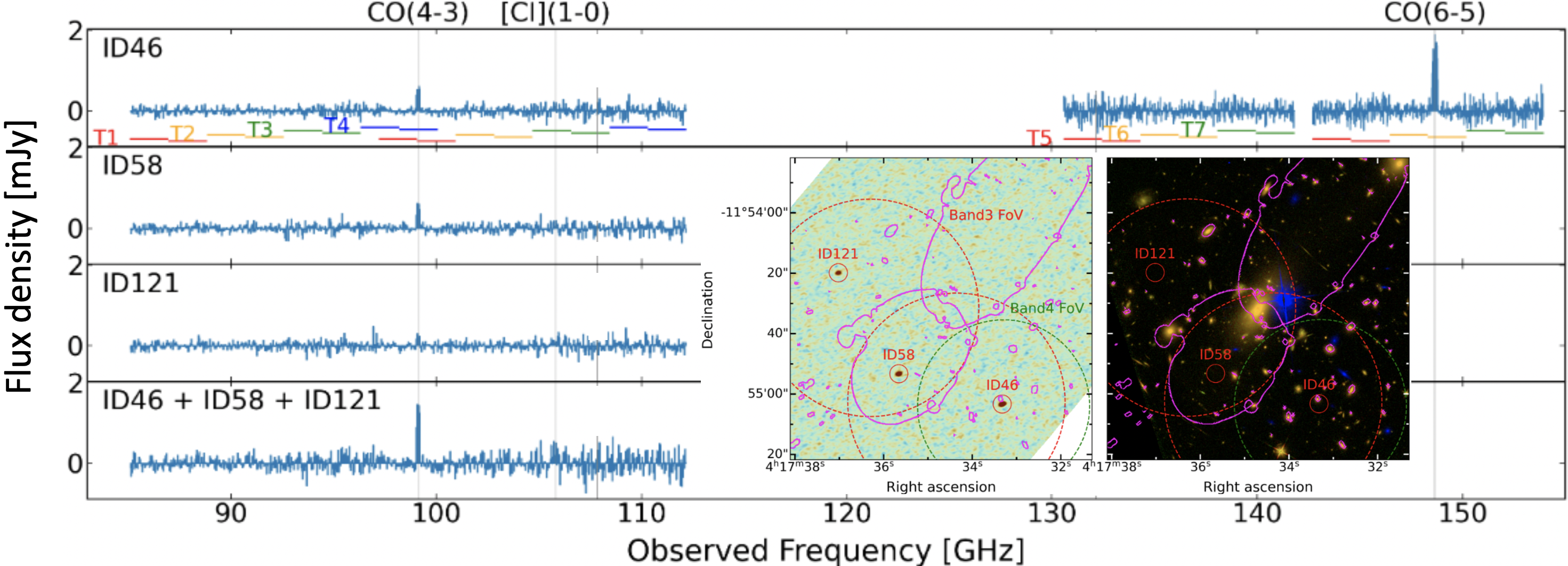}
\caption{ALMA band-3/4 spectra for ALCS 1.2-mm continuum sources, ID46, ID58, and ID121, in MACS J0417.5-1154 produced by combining seven independent tunings (T1--T7). The inserted panels display the ALCS 1.2-mm continuum (left) and \textit{HST} 3-color composite (right; RGB = F160W, F125W, F814W), along with the positions of three sources and the critical curve at $z=3.7$ (magenta contours). 
The thin holizontal lines indicate the expected observed frequencies of the CO(4--3), \CI($^3$P$_1$--$^3$P$_0$), and CO(6--5) at the resultant spectroscopic redshift $z_{\mathrm CO}=3.652$, which is consistent with the predictions from the mass/lens model of this field.}
\label{Kohno:fig:M0417}
\end{figure}

Multiple ALMA follow-up programs to conduct spectral scans toward \textit{H}- and IRAC-dropout ALCS 1.2 mm sources have been conducted. Figure~\ref{Kohno:fig:M0417} shows a triple image \textit{H}-dropout ALMA source behind MACS J0417.5-1154 and its band-3/4 spectra, giving  $z_{\rm CO} = 3.652$ (Tsujita et al., in prep.). 

\section{Number Counts}

\begin{figure}[htp]
\begin{center}
\includegraphics[angle=0.0,width=0.8\textwidth]{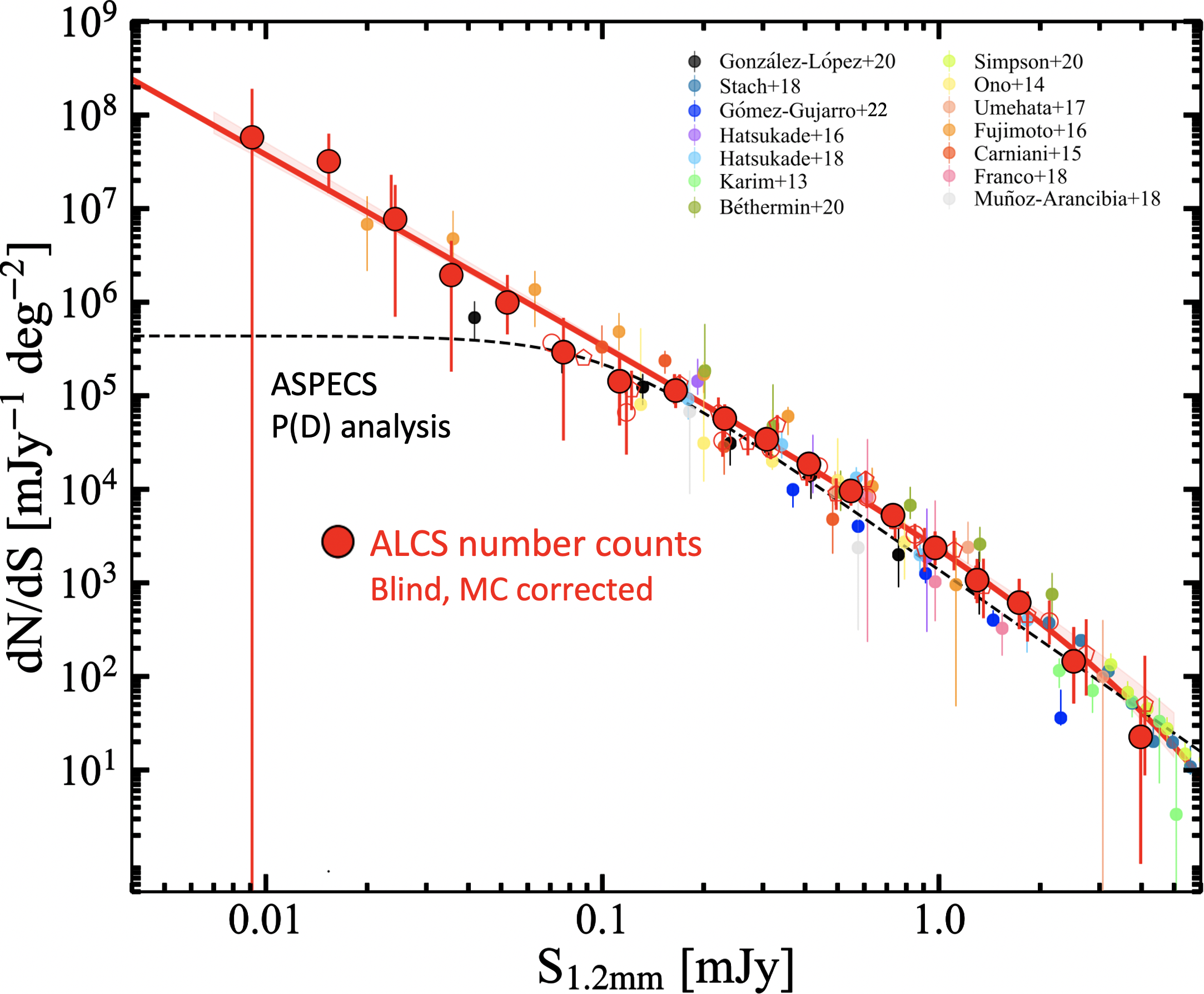}
\end{center}
\caption{
Differential number counts at 1.2 mm \citep{Fujimoto2023}. 
The filled red circles represent the ALCS number counts using the blindly identified sources corrected by the MC simulations that implement relevant uncertainties such as $z$, $\mu$, and flux density measurements, where the error bars indicate the 16-84th percentile in the 5,000 MC realizations. The red line and the shaded region denote the best-fit Schechter function and the associated 1$\sigma$ error.
The faint-end flattening of the ASPECS number counts estimated by \textit{P(D)} analysis \citep{Gonzalez-Lopez2020} is shown (dashed line). 
For measurements observed at different wavelengths from 1.2 mm, we scale the flux density by assuming a typical FIR SED shape based on a single modified black body with $T_{\rm dust}$ = 35 K, spectral index $\beta$ = 1.8, and $z$ = 2.
}
\label{Kohno:fig:NumCounts}
\end{figure}

Figure~\ref{Kohno:fig:NumCounts} shows the differential number counts constructed by the ALCS 1.2 mm continuum source catalog. A detailed description of number counts and subsequent discussions will be given in a forthcoming paper \citep{Fujimoto2023}. With the blind sample, we derive 1.2-mm number counts down to $\sim$7 $\mu$Jy assisted by the gravitational lensing, and find that the total integrated 1.2 mm flux of the securely identified sources corresponds to $\sim$80\% of the CIB light. We find a steady increase 
in the number counts even below $S_{\rm 1.2 mm}<0.1-0.01$ mJy, which 
does not support the flattening shape argued 
in HUDF \citep{Gonzalez-Lopez2020}. It urges us to 
further investigate the faint-end of the 1-mm number counts by extending 
unbiased ALMA deep surveys 
and/or systematically increasing the spec-$z$ sample in the lensing fields
and ultimately improving the precision of the amplification estimates.

\section{Synergies with JWST and future prospects}

An exceptionally magnified ($\mu \sim$20--160), intrinsically faint (sub-$L^{*}$-class) low-mass ($M_{\rm *}\sim10^9$ $M_\odot$) star-forming galaxy at $z=6.07$ \citep{Laporte2021,Fujimoto2021}, which is one of the early outcomes of an unbiased line emitter search using the ALCS 3D cube, is scheduled to be observed by \textit{JWST}/NIRCam, NIRSpec, VLT/MUSE, and so on. \textit{JWST} observations of lensing clusters including ALCS fields are now rapidly increasing, allowing us to characterize ALCS 1.2 mm continuum sources in e.g., SMACS J0723.3-7327 \citep{Cheng2022,Fudamoto2022} 
and Abell 2744 \citep{Kokorev2023}. 

Recent \textit{JWST}/NIRCam observations report a remarkably high abundance of NIR-dropout sources, i.e., ``ultra-high-redshft'' LBG candidates (e.g., \citealt{Finkelstein2023,Bouwens2023,Donnan2023,Harikane2023}), which may violate the current galaxy formation models based on the $\Lambda$-CDM framework \citep{Naidu2022,Lovell2023}. ALMA spectroscopy observations of an ultra-high-$z$ candidate suggest that a part of such ultra-high-$z$ candidates may be dust-enshrouded star-forming galaxies at $z\sim4-5$
\citep{Fujimoto2022}, implying a potential overlap with the \textit{H}-band dropout sources. Therefore, further investigation of the physical properties of \textit{H}- and IRAC-dropout sources uncovered by ALCS will help the interpretation of the putative ultra-high-redshift candidates.

%


\begin{thebibliography}{99}
\bibitem[Aravena et al.(2016)]{Aravena2016} Aravena, M.; Decarli, R.; Walter, F. et al.\ 2016, ApJ, 833, 68
\bibitem[Aravena et al.(2020)]{Aravena2020} Aravena, M.; Boogaard, L.; G{\'o}nzalez-L{\'o}pez, J.\ et al.\ 2020, ApJ, 901, 79
\bibitem[B{\'e}thermin et al.(2020)]{Bethermin2020} B{\'e}thermin, M.; Fudamoto, Y.; Ginolfi, M.\ et al.\ 2020, A\&A, 643, A2
\bibitem[Bouwens et al.(2023)]{Bouwens2023} Bouwens, R., Illingworth, G., Oesch, P., et al.\ 2023, MNRAS, in press
\bibitem[Carniani et al.(2015)]{Carniani2015} Carniani, S.; Maiolino, R.; De Zotti, G.\ et al.\ 2015, A\&A, 584, A78
\bibitem[Caputi et al.(2014)]{Caputi2014} Caputi, K.~I.; Micha{\l}owski, M.~J.; Krips, M.\ et al.\ 2014, ApJ, 788, 126
\bibitem[Caputi et al.(2021)]{Caputi2021} Caputi, K.~I.; Caminha, G.~B.; Fujimoto, S. et al.\ 2021, ApJ, 908, 146
\bibitem[Casey et al.(2019)]{Casey2019} Casey, C.~M.; Zavala, J.~A.; Aravena, M.\ et al.\ 2019, ApJ, 887, 55
\bibitem[Casey et al.(2021)]{Casey2021} Casey, C.~M.; Zavala, J.~A.; Manning, S.~M.\ et al.\ 2021, ApJ, 923, 215
\bibitem[Chen et al.(2023)]{Chen2023} Chen, J.; Ivison, R.~J.; Zwaan, M.~A. et al.\ 2023, MNRAS, 518, 1378
\bibitem[Cheng et al.(2022)]{Cheng2022} Cheng, C.; Yan, H.; Huang, J.-S.\ et al.\ 2022, ApJL, 936, L19
\bibitem[Coe et al.(2019)]{Coe2019} Coe, D.; Salmon, B.; Brada{\v{c}}, M.\ et al.\ 2019, ApJ, 884, 85
\bibitem[Cowie et al.(2018)]{Cowie2018} Cowie, L.~L.; Gonz{\'a}lez-L{\'o}pez, J.; Barger, A.~J.\ et al.\ 2018, ApJ, 865, 106
\bibitem[Dessauges-Zavadsky et al.(2017)]{Dessauges-Zavadsky2017} Dessauges-Zavadsky, M.; Zamojski, M.; Rujopakarn, W.\ et al.\ 2017, A\&A, 605, A81
\bibitem[Donnan et al.(2023)]{Donnan2023} Donnan, C.~T., McLeod, D.~J., Dunlop, J.~S., et al.\ 2023, MNRAS, 518, 6011
\bibitem[Dunlop et al.(2017)]{Dunlop2017} Dunlop, J.~S.; McLure, R.~J.; Biggs, A.~D., et al.\ 2017, MNRAS, 466, 861
\bibitem[Finkelstein et al.(2023)]{Finkelstein2023} Finkelstein, S.~L., Bagley, M.~B., Ferguson, H.~C., et al.\ 2023, ApJL, 946, L13
\bibitem[Franco et al.(2018)]{Franco2018} Franco, M.; Elbaz, D.; B{\'e}thermin, M. et al.\ 2018, A\&A, 620, A152
\bibitem[Fudamoto et al.(2022)]{Fudamoto2022} Fudamoto, Y.; Inoue, A.~K.; \& Sugahara, Y.\ 2022, ApJL, 938, L24
\bibitem[Fujimoto et al.(2016)]{Fujimoto2016} Fujimoto, S.; Ouchi, M.; Ono, Y. et al.\ 2016, ApJS, 222, 1
\bibitem[Fujimoto et al.(2021)]{Fujimoto2021} Fujimoto, S.; Oguri, M.; Brammer, G.\ et al.\ 2021, ApJ, 911, 99
\bibitem[Fujimoto et al.(2022)]{Fujimoto2022} Fujimoto, S.;  Finkelstein, S.~L.; Burgarella, D., et al.\ submitted to ApJ (arXiv:2211.03896)
\bibitem[Fujimoto et al.(2023)]{Fujimoto2023} Fujimoto, S., Kohno, K., Ouchi, M., et al.\  submitted to ApJS (arXiv:2303.01658)
\bibitem[G{\'o}mez-Guijarro et al.(2022)]{Gomez-Guijarro2022} G{\'o}mez-Guijarro, C.; Elbaz, D.; Xiao, M.; et al.\ 2022, A\&A, 658, A43
\bibitem[Gonz{\'a}lez-L{\'o}pez et al.(2017)]{Gonzalez-Lopez2017} Gonz{\'a}lez-L{\'o}pez, J.; Bauer, F.~E.; Aravena, M.\ et al.\ 2017, A\&A, 608, A138
\bibitem[Gonz{\'a}lez-L{\'o}pez et al.(2020)]{Gonzalez-Lopez2020} Gonz{\'a}lez-L{\'o}pez, J.; Novak, M.; Decarli, R.\ et al.\ 2020, ApJ, 897, 91
\bibitem[Grogin et al.(2011)]{Grogin2011} Grogin, N.~A., Kocevski, D.~D., Faber, S.~M., et al.\ 2011, ApJS, 197, 35
\bibitem[Gruppioni et al.(2020)]{Gruppioni2020} Gruppioni, C.; B{\'e}thermin, M.; Loiacono, F.\ et al.\ 2020, A\&A, 643, A8
\bibitem[Harikane et al.(2023)]{Harikane2023} Harikane, Y.; Ouchi, M., Oguri, M., et al.\ ApJS, in press (arXiv:2208.01612)
\bibitem[Hatsukade et al.(2013)]{Hatsukade2013} Hatsukade, B.; Ohta, K.; Seko, A.\ et al.\ 2013, ApJL, 769, L27
\bibitem[Hatsukade et al.(2016)]{Hatsukade2016} Hatsukade, B.; Kohno, K.; Umehata, H.\ et al.\ 2016, PASJ, 68, 36
\bibitem[Hatsukade et al.(2018)]{Hatsukade2018} Hatsukade, B.; Kohno, K.; Yamaguchi, Y.\ et al.\ 2018, PASJ, 70, 105
\bibitem[Jolly et al.(2021)]{Jolly2021} Jolly, J.-B.; Knudsen, K.; Laporte, N.\ et al.\ 2021, A\&A, 652, A128
\bibitem[Karim et al.(2013)]{Karim2013} Karim, A.; Swinbank, A.~M.; Hodge, J.~A.\ et al.\ 2013, MNRAS, 432, 2
\bibitem[Koekemoer et al.(2011)]{Koekemoer2011} Koekemoer, A.~M., Faber, S.~M., Ferguson, H.~C., et al.\ 2011, ApJS, 197, 36
\bibitem[Kokorev et al.(2022)]{Kokorev2022} Kokorev, V.; Brammer, G.; Fujimoto, S.\ et al.\ 2022, ApJS, 263, 38
\bibitem[Kokorev et al.(2023)]{Kokorev2023} Kokorev, V.; Jin, S.; Magdis, G.~E. et al.\ submitted to ApJL (arXiv:2301.04158)
\bibitem[Knudsen et al.(2008)]{Knudsen2008} Knudsen, K.~K.; van der Werf, P.~P.; \& Kneib, J.-P.\ 2008, MNRAS, 384, 1611
\bibitem[Lagos et al.(2020)]{Lagos2020} Lagos, C. del P.; da Cunha, E.; Robotham, A.~S.~G.\ et al.\ 2020, MNRAS, 499, 1948
\bibitem[Laporte et al.(2021)]{Laporte2021} Laporte, N.; Zitrin, A.; Ellis, R.~S.\ et al.\ 2021, MNRAS, 505, 4838
\bibitem[Lotz et al.(2017)]{Lotz2017} Lotz, J. M.; Koekemoer, A.; Coe, D.\ et al.\ 2017, ApJ, 837, 97
\bibitem[Lovell et al.(2023)]{Lovell2023} Lovell, C.~C., Harrison, I., Harikane, Y., et al.\ 2023, MNRAS, 518, 2511
\bibitem[Manning et al.(2022)]{Manning2022} Manning, S.~M.; Casey, C.~M.; Zavala, J.~A.\ et al.\ 2022, ApJ, 925, 23
\bibitem[Mu{\~n}oz Arancibia et al.(2018)]{MunozArancibia2018} Mu{\~n}oz Arancibia, A.~M.; Gonz{\'a}lez-L{\'o}pez, J.; Ibar, E.\ et al.\ 2018, A\&A, 620, A125
\bibitem[Mu{\~n}oz Arancibia et al.(2022)]{MunozArancibia2022} Mu{\~n}oz Arancibia, A.~M.; Gonz{\'a}lez-L{\'o}pez, J.; Ibar, E.\ et al.\ submitted to A\&A (arXiv:2203.06195)
\bibitem[Naidu et al.(2022)]{Naidu2022} Naidu, R.~P., Oesch, P.~A., Setton, D.~J., et al.\ submitted to ApJL (arXiv:2208.02794)
\bibitem[Ono et al.(2014)]{Ono2014} Ono, Y.; Ouchi, M.; Kurono, Y.; et al.\ 2014, ApJ, 795, 5
\bibitem[Postman et al.(2012)]{Postman2012} Postman, M.; Coe, D.; Ben{\'i}tez, N.\ et al.\ 2012, ApJS, 199, 25
\bibitem[Schreiber et al.(2018)]{Schreiber2018} Schreiber, C.; Labb{\'e}, I.; Glazebrook, K.\ et al.\ 2018, A\&A, 611, A22
\bibitem[Shu et al.(2022)]{Shu2022} Shu, X.; Yang, L.; Liu, D.\ et al.\ 2022, ApJ, 926, 155
\bibitem[Simpson et al.(2014)]{Simpson2014} Simpson, J.~M.; Swinbank, A.~M.; Smail, I.\ et al.\ 2014, ApJ, 788, 125
\bibitem[Simpson et al.(2020)]{Simpson2020} Simpson, J.~M.; Smail, I.; Dudzevi{\v{c}}i{\={u}}t{\.{e}}, U.\ et al.\ 2020, MNRAS, 495, 3409
\bibitem[Smail et al.(1997)]{Smail1997} Smail, I.; Ivison, R.~J.; \& Blain, A.~W.\ 1997, ApJL, 490, L5
\bibitem[Smail et al.(2021)]{Smail2021} Smail, I.; Dudzevi{\v{c}}i{\={u}}t{\.{e}}, U.; Stach, S.~M.\ et al.\ 2021, MNRAS, 502, 3426
\bibitem[Stach et al.(2018)]{Stach2018} Stach, S.~M.; Smail, I.; Swinbank, A.~M.\ et al.\ 2018, ApJ, 860, 161
\bibitem[Sun et al.(2022)]{Sun2022} Sun, F.; Egami, E.; Fujimoto, S.; et al.\ 2022, ApJ, 932, 77
\bibitem[Sun et al.(2021)]{Sun2021} Sun, F.; Egami, E.; P{\'e}rez-Gonz{\'a}lez, P.~G.\ et al.\ 2021, ApJ, 922, 114
\bibitem[Uematsu et al.(2023)]{Uematsu2023} Uematsu, R.; Ueda, Y.; Kohno, K.\ et al.\ ApJ, in press (arXiv:2301.09275)
\bibitem[Umehata et al.(2017)]{Umehata2017} Umehata, H.; Tamura, Y.; Kohno, K.\ et al.\ 2017, ApJ, 835, 98
\bibitem[Umehata et al.(2020)]{Umehata2020} Umehata, H.; Smail, I.; Swinbank, A.~M.\ et al.\ 2020, A\&A, 640, L8
\bibitem[Wang et al.(2019)]{Wang2019} Wang, T.; Schreiber, C.; Elbaz, D.\ et al.\ 2019, Nature, 572, 211
\bibitem[Williams et al.(2019)]{Williams2019} Williams, C.~C.; Labbe, I.; Spilker, J.\ et al.\ 2019, ApJ, 884, 154
\bibitem[Yamaguchi et al.(2019)]{Yamaguchi2019} Yamaguchi, Y.; Kohno, K.; Hatsukade, B.\ et al.\ 2019, ApJ, 878, 73

\end{thebibliography}
\end{document}